\documentclass[%
reprint,
amsmath,amssymb,
aps,
prb,
]{revtex4-2}

\usepackage{graphicx}
\usepackage{dcolumn}
\usepackage{bm}

\begin{document}

	\title{Unveiling the Impact of Sulfur Doping on Copper-Substituted Lead Apatite: A
		Theoretical Study}
	
	\author{Ming-Long Wang}

	\affiliation{School of Physics and Optoelectronics, South China University of Technology, Guangzhou 510640, People’s Republic of China}
	
	\author{Yin-Hui Peng}
	\affiliation{School of Physics and Optoelectronics, South China University of Technology, Guangzhou 510640, People’s Republic of China}
	
	\author{Ji-Hai Liao}
	\affiliation{School of Physics and Optoelectronics, South China University of Technology, Guangzhou 510640, People’s Republic of China}
	
	\author{Xiao-Bao Yang}
	\email[]{scxbyang@scut.edu.cn}
	\affiliation{School of Physics and Optoelectronics, South China University of Technology, Guangzhou 510640, People’s Republic of China}
	
	\author{Yao Yao}
	\email[]{yaoyao2016@scut.edu.cn}
	\affiliation{School of Physics and Optoelectronics, South China University of Technology, Guangzhou 510640, People’s Republic of China}
	
	\author{Yu-Jun Zhao}
	\email[]{zhaoyj@scut.edu.cn}
	\affiliation{School of Physics and Optoelectronics, South China University of Technology, Guangzhou 510640, People’s Republic of China}
	
	
\begin{abstract}
	Room-temperature superconductivity represents a significant scientific milestone, with the initial report of LK-99, a copper-substituted lead apatite $\mathrm{Pb}_{10-x}\mathrm{Cu}_{x}(\mathrm{PO}_{4})_{6}\mathrm{O}$, offering a potential breakthrough. However, other researchers have encountered numerous challenges in replicating the original experimental results. In recent studies, Wang et al. successfully observed signs of a possible superconducting phase, such as smaller resistance and stronger diamagnetism, upon doping S into the samples. This indicates that the introduction of S is of significant importance for achieving an appropriate structure. To further investigate the role of S, we have considered the $\mathrm{Pb}_{10-x}\mathrm{Cu}_{x}(\mathrm{PO}_{4})_{6}\mathrm{S}$, systematically discussing its thermodynamic stability, as well as the influence of S on the distribution, concentration, and electronic properties of Cu. We find that $\mathrm{Pb}_{10-x}\mathrm{Cu}_{x}(\mathrm{PO}_{4})_{6}\mathrm{S}$ maintains thermodynamic stability, with S primarily influencing the distribution of Cu. The critical element	dictating the electronic characteristics of the	material post-synthesis is Cu, while the impact of S on the electronic properties is relatively minor. Our work provides valuable insights into the synthesis of potential apatite based room-temperature superconductors and the role of S in facilitating Cu doping.

\end{abstract}
	
	\maketitle

\section{INTRODUCTION}
	
	The pursuit of room-temperature superconductivity has long been a holy grail in the field of condensed matter physics, owing to its revolutionary potential in transforming energy transmission, magnetic levitation, and quantum computing technologies. The report of LK-99, a copper-substituted lead apatite $\mathrm{Pb}_{10-x}\mathrm{Cu}_{x}(\mathrm{PO}_{4})_{6}\mathrm{O}$ (PCPO), claimed as a promising candidate for room-temperature superconductivity\cite{lee2023first,lee2023superconductor}, has sparked considerable debate and controversy within the scientific community.
	
	However, other researchers have encountered numerous challenges in replicating the original experimental results, with the synthesized samples failing to demonstrate any superconducting phenomena \cite{puphal2023single,liu2023semiconducting,wang2023ferromagnetic,kumar2023absence,zhu2023firstorder,timokhin2023synthesis,zhang2024ferromagnetism,thakur2023synthesis,lei2024characteristics}. Observations during experimental synthesis have revealed that Cu is difficult to dope into the parent compound $\mathrm{Pb}_{10}(\mathrm{PO}_{4})_{6}\mathrm{O}$ (PPO) \cite{zhang2024ferromagnetism,thakur2023synthesis,puphal2023single}. Theoretical calculations have also confirmed this, revealing that concentration of Cu doping is only 0.24\% \cite{toriyama2024defect}, which does not align with the experimental requirement of a $5\sim10$\% concentration \cite{lee2023first,lee2023superconductor}. Additionally, the synthesis process may result in the formation of the byproduct Cu$_{2}$S, suggested to be the source of the electrical behavior observed in the original documentation of LK-99\cite{zhu2023firstorder,jain2023superionica,lei2024characteristics,liu2023phases}. To circumvent the interference caused by Cu$_{2}$S, Puphal et al. using the traveling solvent floating zone growth method to synthesize single crystals of $\mathrm{Pb}_{10-x}\mathrm{Cu}_{x}(\mathrm{PO}_{4})_{6}\mathrm{O} \ (x=1)$, which are highly insulating and optically transparent, leading to the conclusion that these crystals are most likely not superconducting \cite{puphal2023single}. Beyond this, Density Functional Theory (DFT) calculations have found that it is only when Cu substitutes for the Pb(1) sites that LK-99 exhibits band crossing at the Fermi level\cite{yang2023initio,sun2023metallization,korotin2023electronic,si2023electronic,liu2023symmetry}. However, the energy for Cu substitution at the two distinct crystallographic sites in the parent compound of LK-99, denoted as Pb(1) and Pb(2), is very close, with a slight preference for substituting the Pb(2) site \cite{yang2023initio, sun2023metallization,liu2023symmetry}. This implies that, under the already challenging conditions for Cu doping into the parent compound, the randomness in the available substitution sites for Cu further reduces the probability of achieving an appropriate structure for superconductivity.
	
	Recent articles have reported observing possible signs of room-temperature superconductivity upon doping S into PPO, including the observation of a possible Meissner effect and a reduction in room-temperature resistivity from an insulating state to $2 \times 10^{-5} \  \Omega \cdot \mathrm{m}$\cite{wang2024possible,wang2024observation}. This suggests that the introduction of S into the material may affect aspects such as Cu doping concentration, defect configuration, and electronic properties. However, the available calculations have been performed on the PPO parent compound without considering the sulfur-containing parent compound. Therefore, a systematic study is necessary. Considering the stability of the PO$_4$ group, it may be challenging for S doping to occur within it. Previous theoretical calculations have confirmed that in a synthesis environment with excess S, S is likely to replace oxygen atoms at the center of Pb(2) site\cite{toriyama2024defect}. And the density of states (DOS) indicates that the contribution to the flat band primarily originates from Cu and O at the center of the Pb(2) sites\cite{korotin2023electronic, si2023electronic, tao2024occupied}. To simplify the analysis, we will only discuss the structure where S replaces the oxygen atom at the Pb(2) center in $\mathrm{Pb}_{10}(\mathrm{PO}_{4})_{6}\mathrm{S}$ (PPOS).
	
	Here, we briefly discuss the difference of structure between the parent compound PPO and its sulfur-containing variant, PPOS. We present the calculated phase diagrams for Pb-P-O and Pb-P-O-S, with Cu-S-O constraints applied to the latter, to establish the chemical potential ranges for PPO and PPOS. Subsequently, the impact of S introduction on the preference for Cu substitution sites is discussed, as well as the influence between different channel and the concentration of Cu. Moreover, we carefully analyze the properties of the most stable structure we found. Interestingly, our findings reveal that Cu preferentially substitutes the Pb(2) sites, with the Cu defects tending towards a disordered distribution. Although the stable configurations identified exhibit semiconductor properties, the alignment with experimental synthesis procedures provides valuable direction for the discovery of an appropriate structure for possible apatite based superconductivity.

\section{COMPUTATIONAL DETAILS}
	
	We use Structures of Alloy Generation And Recognition (SAGAR)\cite{he2020atom,he2021biased} to generate the nonequivalent substitution structures of Cu in different distribution. The subsequent calculations are performed in the framework of the Density Functional Theory (DFT) as implemented in the Vienna \textit{ab initio} Simulation Package (VASP)\cite{kresse1996efficiency,kresse1996efficient}. The PerdewBurke–Ernzerhof (PBE) functional\cite{perdew1992atoms} in Generalized Gradient Approximation (GGA)\cite{perdew1996generalized} was set to describe the exchange-correlation energy. The plane-wave cutoff was chosen to be 500 eV. A $\Gamma$-centered $k$-point grid with a $k$-spacing of 0.05 $2\pi /\mathring{\mathrm{A}}$ was used for structural relaxations, the convergence criteria for the total energy and ionic forces were set to $10^{-5}$ eV and 0.01 eV$/\mathring{\mathrm{A}}$, respectively. In the self-consistent field (SCF) calculations, as well as in the computations of band structures and density of states (DOS), we reduce the $k$-spacing to 0.03 $2\pi /\mathring{\mathrm{A}}$, and $10^{-8}$ eV for the convergence criteria. The electronic correlation effects were described using the $\mathrm{DFT}+U$ method based on the Dudarev scheme, $U=4$ eV\cite{dudarev1998electronenergyloss}.
	
\section{THE MODIFIED STRUCTURE OF LEAD APATITE}
	
	\begin{figure*}[htpb]
		\includegraphics[width=0.8\textwidth]{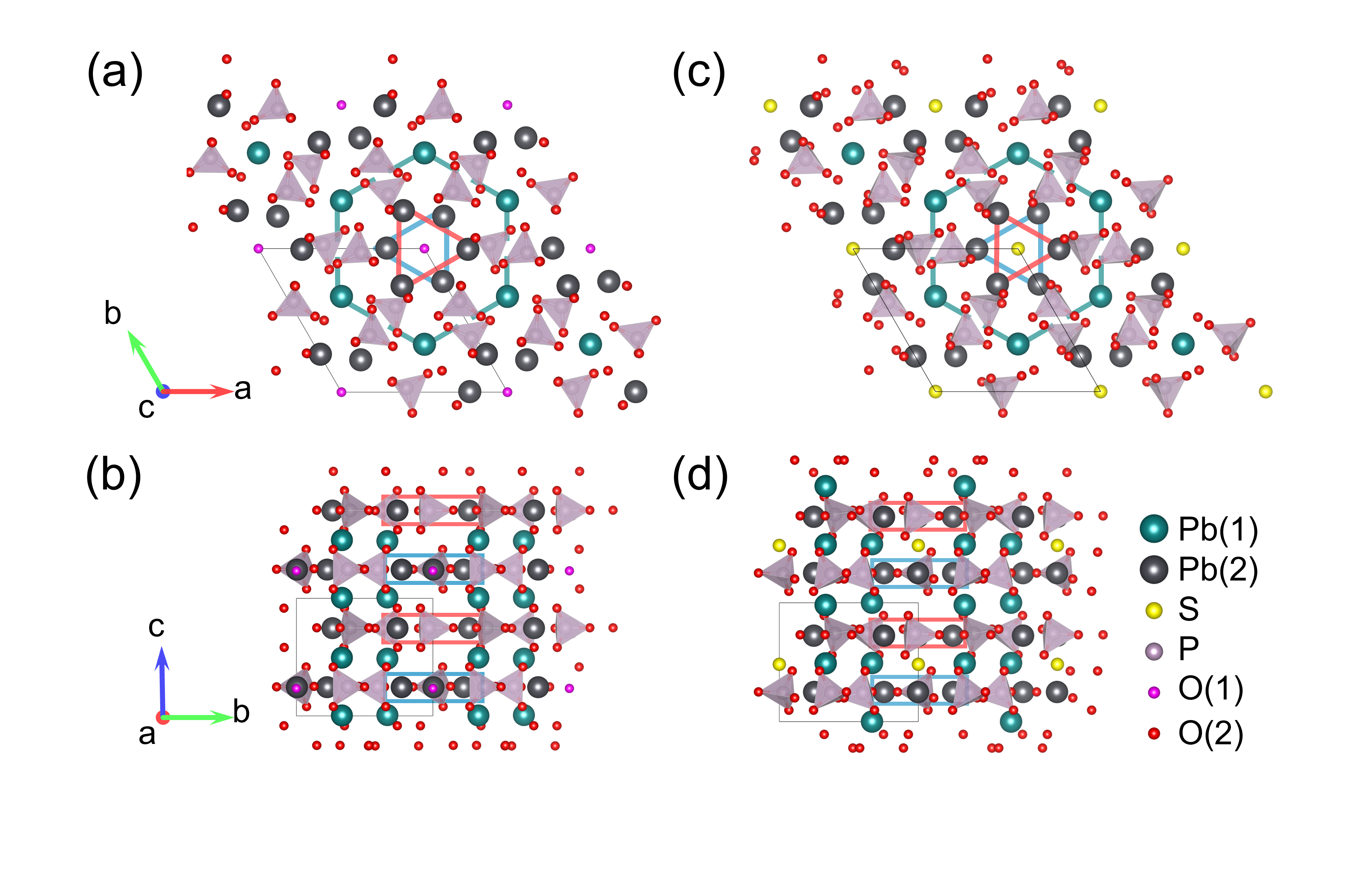}
		\caption{\label{fig1} Top (a) and side (b) views of the parent compound of LK-99 with chemical formula $\mathrm{Pb}_{10}(\mathrm{PO}_{4})_{6}\mathrm{O}$ (PPO). The green lines indicate the channel formed by Pb(1) sites, while the blue and red lines denote the oppositely shaped triangle form by Pb(2) sites. Top (c) and side (d) views of the crystal structure of PPOS with chemical formula $\mathrm{Pb}_{10}(\mathrm{PO}_{4})_{6}\mathrm{S}$ (PPOS).}
	\end{figure*}

	The parent compound of LK-99 is lead-phosphate with chemical formula $\mathrm{Pb}_{10}(\mathrm{PO}_{4})_{6}\mathrm{O}$ (PPO)\cite{lee2023first,lee2023superconductor}. In PPO, as shown in Fig.~\ref{fig1}, there exists two symmetry nonequivalent Pb sites, named Pb(1) and Pb(2). The Pb(1) sites form a one-dimensional channel along the $c$-axis, as indicated by the green lines in Fig.~\ref{fig1}. The Pb(2) sites are clustered into oppositely shaped triangular configurations across two layers, represented by blue and red solid lines in Fig.~\ref{fig1}. An oxygen atom which is situated at the center of the channel and lies co-planar with the triangles formed by the Pb(2) sites. Defect formation energies for Cu doping into PPO, either the Pb(1) or Pb(2) sites are very similar\cite{yang2023initio, sun2023metallization,liu2023symmetry}, suggesting a high degree of randomness in the substitution sites during experimental synthesis. This randomness could account for the difficulty many studies faced in reproducing samples that exhibits levitation.
	
	In a subsequent experimental synthesis by Wang et al.\cite{wang2024possible}, the replacement of the central O atom at the Pb(2) site with an S atom resulted in samples that consistently demonstrated the Messiner effect. This suggests that the introduction of S atoms may significantly influence the doping of Cu in many aspects. Considering the PO$_4$ group's stability and prior DFT calculations\cite{toriyama2024defect}, we will solely consider the configuration where S substitutes for the central O atom at the Pb(2) sites, aligning with the formula $\mathrm{Pb}_{10-x}\mathrm{Cu}_{x}(\mathrm{PO}_{4})_{6}\mathrm{S}$ (PPOS) as presented by Wang et al.\cite{wang2024possible}.
	
	Replacing the central O atom at the Pb(2) sites in PPO with an S atom. After structure relaxation, S migrate to the center of the interlayer space between the two triangular planar. This migration of S slightly increases the lattice constants along the $ab$ plane and a decrease along the $c$-axis, indicating an adjustment of the primitive cell dimensions to accommodate the larger S atoms while maintaining the overall structural integrity. Additionally, in Fig.\ref{fig1}(c,d), due to the presence of S vacancies, we can consider a layer consisting of S and the set of atoms forming triangles constituted by the adjacent Pb(2) sites.
	
	\begin{table}[htpb]
	\centering
	\caption{The lattice parameter of compounds of PPO and PPOS.}
	\begin{tabular}{|c|c|c|c|c|}
		\hline
		 & $a \ \mathring{\mathrm{A}}$ & $c \ \mathring{\mathrm{A}}$ & error in $a$ & error in $c$ \\
		\hline
		Experiment\cite{lee2023first,lee2023superconductor} & 9.865 & 7.431 & & \\
		\hline
		PPO & 10.021 &7.485 & 1.5\% & 0.7\%\\  
		\hline 
		PPOS & 10.032 & 7.439 & 1.6\% & 0.1\% \\
		\hline
	\end{tabular}
	\label{tab1}
	\end{table}

\section{THERMODYNAMIC STABILITY OF PPO AND PPOS}

	\begin{figure*}[htpb]
		\includegraphics[width=0.8\textwidth]{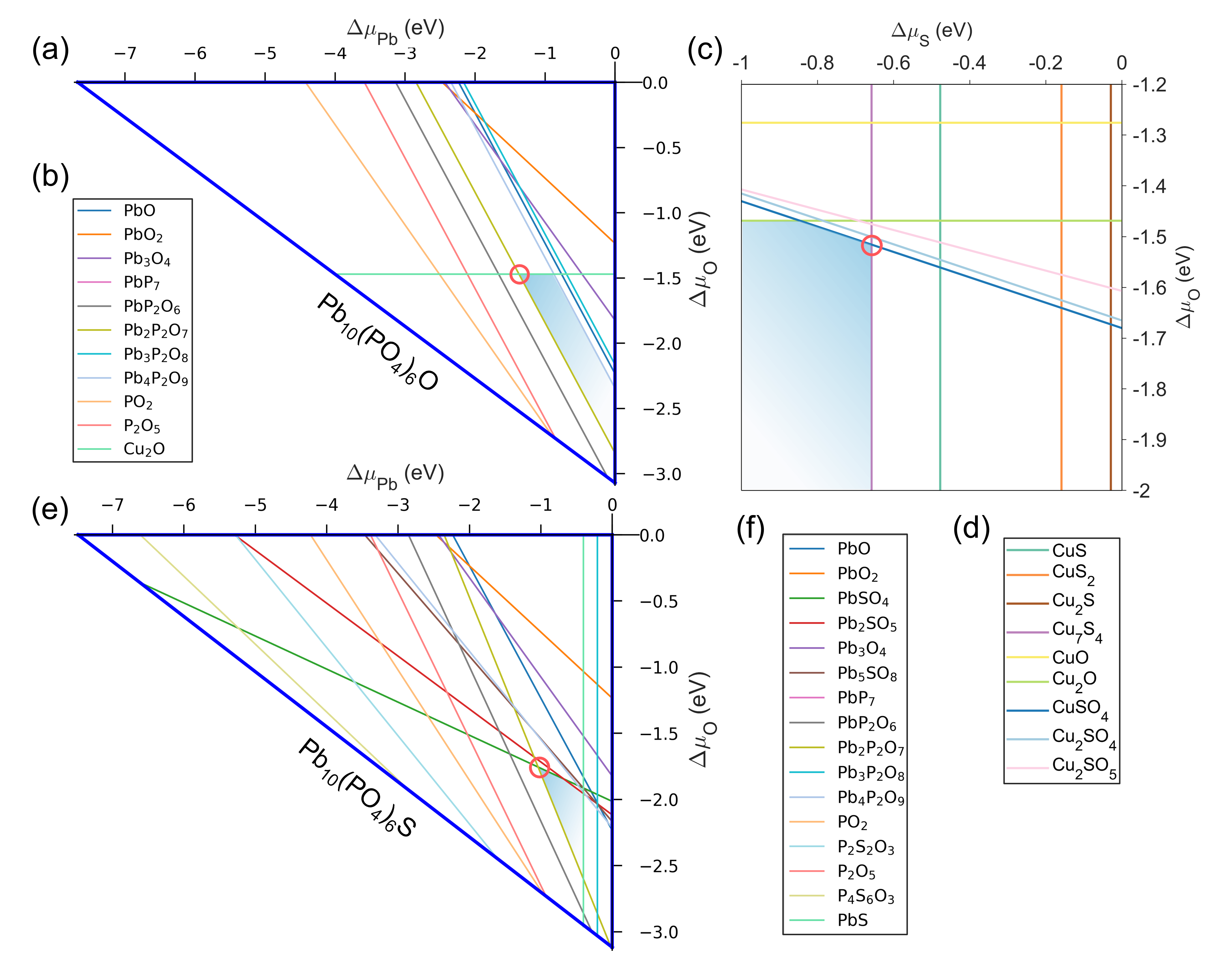}
		\caption{\label{fig2} (a,b) The phase diagram of $\mathrm{Pb}_{10}(\mathrm{PO}_{4})_{6}\mathrm{O}$ projected in $(\Delta\mu_{\mathrm{Pb}},\Delta\mu_{\mathrm{O}})$ plane, where $\Delta\mu_{\mathrm{Cu-rich}}=0 \ \mathrm{eV}$. The blue-shaded area represents the  chemical potential range that the host exists under equilibrium. (c,d) The chemical potential range of S and O in Cu maximum rich condition ($\Delta\mu_{\mathrm{Cu}}=0$). (e,f) The phase diagram of $\mathrm{Pb}_{10}(\mathrm{PO}_{4})_{6}\mathrm{S}$ projected in $(\Delta\mu_{\mathrm{Pb}},\Delta\mu_{\mathrm{O}})$ plane, where $\Delta\mu_{\mathrm{S-rich}}=-0.66 \ \mathrm{eV}$. The red circles in (a,c,e) indicate the chemical potential values where the defect formation energy for Cu substituting Pb is minimized.
		}
	\end{figure*}

	We utilized the defect formation energy to evaluate the difficulty of doping Cu into the host material, with the formula as shown below.
	$$
	\Delta H_{f}^{\alpha, q} = E(\alpha) - E(0) + \sum_{\alpha} n_{\alpha} (\Delta\mu_{\alpha} + \mu_{\alpha}^{\mathrm{solid}}) + q (E_{\mathrm{VBM}} + E_{\mathrm{F}}),
	$$
	where $E(\alpha)$ and $E(0)$ are the total energy of the supercell with and without defect $\alpha$. $n_\alpha$ is the number of each defect atom; $n_\alpha = -1$ if an atom is added, while $n_\alpha=1$ if an atom is removed. $\mu_{\alpha}^{\mathrm{solid}} = -E_{\mathrm{coh}}(\mathrm{Solid})$, $-E_{\mathrm{coh}}(\mathrm{Solid})$ stands for the cohesive energy. $\Delta\mu_{\alpha}$ is the chemical potential with respect to its corresponding elemental crystal. Typically, $\Delta\mu_{\alpha}=0 \ \mathrm{eV}$ corresponds to a maximum $\alpha$-rich condition. $q$ represent the charge state, while $E_{\mathrm{VBM}}$ and $E_{\mathrm{F}}$ represents the energy at the valence band maximum of the defect free system and Fermi energy relative to the $E_{\mathrm{VBM}}$, respectively. As the previous studies have shown that Cu substitution is limited below 0.24\%, much less than 5-10\% range claimed from experiment\cite{lee2023first,lee2023superconductor,wang2024possible,wang2024observation}, we assume that the synthesis is in the maximum Cu-rich condition to achieve a higher substitution. Additionally, Cu defects are expected to be neutrally charged, as the sample stays neutral with such a reported high concentration of Cu defects\cite{lee2023first,lee2023superconductor}. Therefore we do not take into account the influence of charge state in subsequent calculations.
	
	To address the thermodynamic stability of the doped host structure and the subsequent chapters discussing the impact of S incorporation on Cu doping concentration, we have computed the phase diagrams for both the Pb-Cu-P-O and Pb-Cu-P-O-S systems. For Pb-Cu-P-O system, we calculated the formation energies of all competing phases that can form from Pb-P-O, taking into account the potential competing phases that may arise from excess Cu-O-P during synthesis. The resulting phase diagram is depicted in Fig.~\ref{fig2}(a,b), where the blue shaded region represents the chemical potential accessible under thermodynamic equilibrium.
	
	The computation of the chemical potential range for the Pb-Cu-P-O-S system is somewhat more intricate than that for PPO due to the increased elemental diversity, necessitating a separate discussion of the competitive phases for each constituent. First, we calculate the competing phase of Cu-S-O and project it in $(\Delta\mu_{\mathrm{S}}, \Delta\mu_{\mathrm{O}})$ plane, as shown in Fig.~\ref{fig2}(c,d). We can see the chemical potential mainly constrained by $\mathrm{CuSO}_4$ and $\mathrm{Cu}_7\mathrm{S}_4$. In order to reduce the defect formation energy of Cu in Pb-Cu-P-O-S system, it is necessary to lower the chemical potential of Pb, corresponding to S in rich condition $\Delta\mu_{\mathrm{S}}=-0.66 \ \mathrm{eV}$. Subsequently, we fix the chemical potential of S and consider the phase diagram of Pb-P-O-S for host PPOS, as shown in Fig.~\ref{fig2}(e,f). The chemical potential is mainly constrained by $\mathrm{PbSO}_4$, $\mathrm{Pb}_2\mathrm{P}_2\mathrm{O}_7$ and PbS. Then we employ the following equation to compute the defect formation energy of Cu substituting Pb:	
	$$
	\begin{aligned}
		\Delta H_{f}^{x}(\mathrm{Cu}_{\mathrm{Pb}}) = 
		E(\mathrm{Pb}_{10-x}\mathrm{Cu}_{x}(\mathrm{PO}_{4})_{6}\mathrm{S}) - E(\mathrm{Pb}_{10}(\mathrm{PO}_{4})_{6}\mathrm{S})  \\
		+ x \left[ \left( \Delta\mu_{\mathrm{Pb}} + \mu_{\mathrm{Pb}}^{\mathrm{solid}}\right)  - \left( \Delta\mu_{\mathrm{Cu}} + \mu_{\mathrm{Cu}}^{\mathrm{solid}}\right) \right] \\
	\end{aligned}
	$$
	
	To minimize the defect formation energy, it is necessary to set the chemical potential of Pb in poor conditions, while Cu in maximum-rich condition. The points that we ultimately select are indicated with red circles in Fig.~\ref{fig2}(a,e). In Pb-Cu-P-O system, we choose $\Delta\mu_{\mathrm{Cu}} = 0$ and $\Delta\mu_{\mathrm{Pb}} = -1.37 \ \mathrm{eV}$. And in Pb-Cu-P-O-S system, we choose $\Delta\mu_{\mathrm{Cu}} = 0$ and $\Delta\mu_{\mathrm{Pb}} = -1.04 \ \mathrm{eV}$.

\section{Configuration of $\bm{\mathrm{Pb}_{10-x}\mathrm{Cu}_{x}(\mathrm{PO}_{4})_{6}\mathrm{O}}$}
	
	As we currently lack information on the defect formation energies for structures with different concentrations of Cu, we initiate our discussion with a concentration consistent with the experiment ($x=1$) as an initial attempt\cite{wang2024possible}. To consider the influence of configuration to the formation energy, we have replaced all possible configuration of two Pb ions with Cu within a $1\times1\times2$ supercell. Utilizing the Structure of Alloy Generation And Recognition (SAGAR)\cite{he2020atom,he2021biased}, we have generated 44 nonequivalent configurations to explore these effects. 
	The final formation energies of the structures are depicted in Fig.~\ref{fig3}(a), where blue circles represent the $1\times1\times2$ supercell. The horizontal axis categorizes the structures based on the Pb sites substituted by Cu. It is evident from the results that the formation energy is significantly higher when Cu substitutes the Pb(1) sites compared to the Pb(2) sites. Previous research on the PCPO structure revealed that the energies for Cu substitution at both Pb(1) and Pb(2) sites are very close\cite{yang2023initio}, which may lead to a high degree of randomness in the experimental synthesis process. This suggests that Cu and S have a strong attraction so that the introduction of S causes Cu to preferentially occupy the Pb(2) sites, thereby increasing the likelihood of obtaining specific structures in experiments. Consequently, we will only consider the scenario where Cu occupies the Pb(2) sites in subsequent discussions.
	
	To discuss whether different distributions of Cu between channels would affect the formation energy, similar to the $1\times1\times2$ supercell, we utilized SAGAR to substitute two Pb atoms with Cu within the $2\times1\times1$ supercell, which possesses two independent channels. The computational results, as indicated by the red diamond symbols in Fig.~\ref{fig3}, show that when Cu is dispersed across two channels, the impact of different Cu distributions on the defect formation energy is relatively small. This suggest the distribution of Cu has no remarkable effect on the defect formation energy between different channels. Consequently, the distribution of Cu between two channels can be regarded as a uniform distribution within a single channel. This is consistent with the stable structure in the $1\times1\times2$ supercell, which also exhibits a uniform distribution. This also explains why, despite a certain degree of computational error between the $1\times1\times2$ and $2\times1\times1$ supercell, the defect formation energies of their most stable structures are very close.
	
	When Cu is aggregated within the same channel, the resulting lowest energy is lower than that of Cu distributed across two channels. Nonetheless, the peak energy values observed can surpass those of the corresponding substitution sites within the $1\times1\times2$ supercell, indicating that the distribution of Cu within a single channel has a significant impact on the defect formation energy.

\section{DEFECT CONCENTRATION OF C$\bm{\mathrm{u}}$}

\begin{figure*}[htpb]
	\centering
	\includegraphics[width=0.8\textwidth]{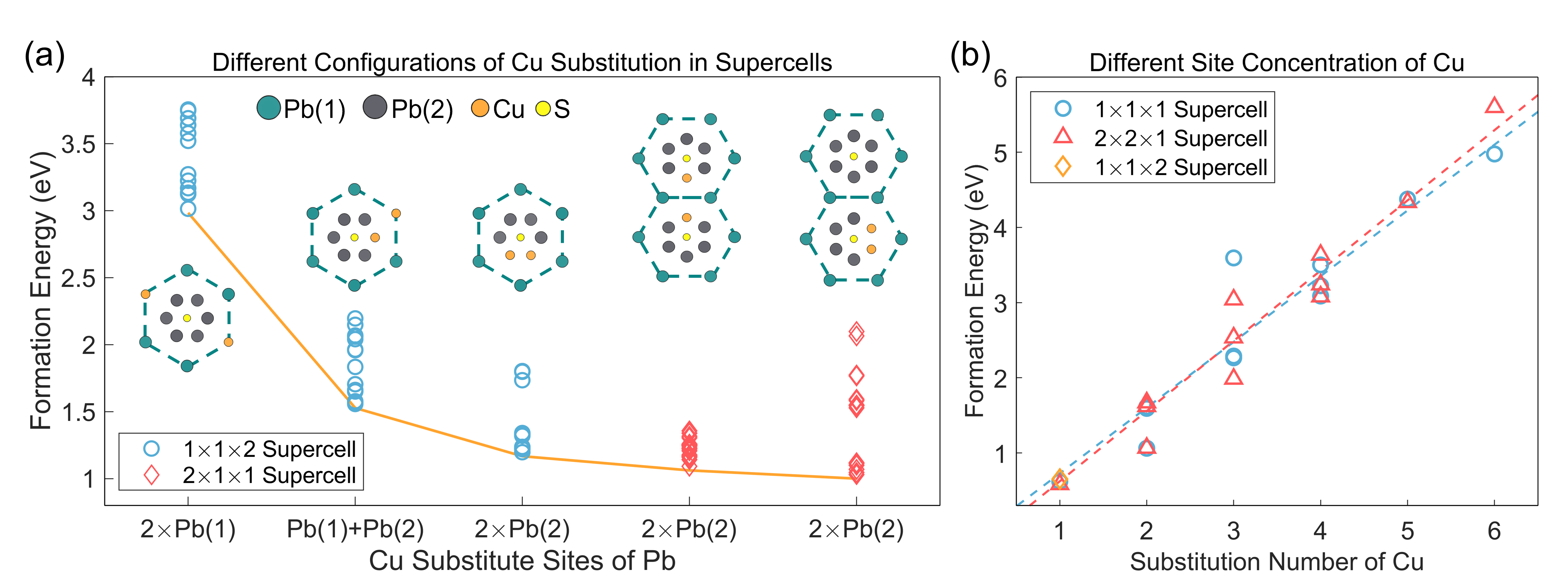} 
	\caption{
		(a) Formation energy of $\mathrm{Pb}_{10-x}\mathrm{Cu}_{x}(\mathrm{PO}_{4})_{6}\mathrm{O}$ with different distributions of Cu. Blue circles represent data from $1\times1\times2$ supercell structure, while red diamonds correspond to $2\times1\times1$ supercell structure. The structures inserted correspond to different columns and are the schematic top-views representation of the structure with the lowest energy. Structures of $2\times1\times1$ supercell structures are classified into two columns based on the number of channels substituted by Cu as shown in the corresponding structure figure. (b) The formation energies for the primitive cell, $1\times1\times2$ and $2\times2\times1$ supercell with varying amounts of Cu substitution and different configuration. Blue circles represent the structures of $1\times1\times2$ supercell, while red triangles and yellow diamond denote the structures of $2\times1\times1$ and $1\times1\times2$ supercell. In the case of $2\times1\times1$ supercell, Cu substitutes only in one of the four independent channels. Dashed lines represent the trend fitted based on the structural formation energies.
		\label{fig3}
	}
\end{figure*}
	
	To explore the optimal doping concentration of Cu within a single channel, we calculated the formation energies for a total of 12 nonequivalent structures with varying numbers of Cu replacing Pb(2) sites in the primitive cell. Furthermore, considering the substantial lattice distortion induced by Cu, we aimed to mitigate the interactions between channels at elevated Cu concentrations. To achieve this, we selectively substituted atoms with Cu in a single channel of the $2\times2\times1$ supercell, which encompasses four distinct channels, while maintaining the equivalence of the structural configurations in this channel with those of the primitive cell.
	
	As shown in Fig.~\ref{fig3}b, the formation energies of the corresponding structures in the $2\times2\times1$ and primitive cells are very close. The energy differences observed in certain structures may arise from the complexity of their potential energy surfaces, leading to entrapment in different local minima during the optimization process. Moreover, structures with significant energy differences are inherently unstable and have relatively higher formation energies. Based on the fitted curves, both datasets exhibit consistent trends, indicating that the variations resulting from structural optimization do not affect subsequent analyses. As the concentration and distribution of Cu across different channels have a negligible impact on the defect formation energy, it implies that the main factor affecting the formation energy is the quantity $x$ of Cu within given channels.
	
	To further investigate whether Cu atoms in the same channel influence each other across different layers, we substituted a single Cu atom into a Pb(2) site within a $1\times1\times2$ supercell, with the formation energy depicted as the orange diamond in Fig.~\ref{fig3}b. It is observed that the formation energy is closely similar to that of the $1\times1\times1$ and $2\times2\times1$ supercell structures, indicating that the distribution of defects within the structure is influenced minimally by interactions not only between channels but also between different layers within a single channel. Thus, we can dissect the structure into layered hexagonal units defined by the channels formed by the Pb(1) sites and the layers constituted by the Pb(2) sites. We propose that this minimal influence is attributed to the interactions between Cu atoms being weaker than those between Cu and S. It is only when multiple Cu atoms are present within the same unit that the mutual interactions between Cu atoms become pronounced, significantly affecting the formation energy. Consequently, with PPOS as the host material, Cu defects are inclined towards a disordered distribution.

	It should be noted that the variable $x$ here quantifies the enrichment of Cu within an unit and does not represent the overall defect concentration. We calculate the site defect concentration $[n]$ at thermal equilibrium using the following formula:
	$$
	[n] = \exp{\left( -\Delta H_{f} / k_{\mathrm{B}} T\right) }, \ \text{(at. \% of all Pb sites)}
	$$
	where $\Delta H_{f}$ is the formation energy we have calculated before and $k_{\mathrm{B}}$ is the Boltzmann constant. Here, the temperature $T$ is assumed to be consistent with the experimental synthesis temperature of 1173.15 K\cite{wang2024possible}. Considering the disorder in defect distribution, we can estimate the thermodynamic equilibrium concentration based on the formation energy of the primitive cell. By selecting the structure with the lowest formation energy at $x=1$, the thermodynamic equilibrium concentration is approximately 0.26\% for the PPOS host, and about 0.10\% for the PPO host. This contrasts with previous studies that concluded a maximum Cu doping concentration of approximately 0.26\% in PPO, which is inconsistent with our findings. This discrepancy may arise from the high sensitivity of the thermodynamic equilibrium concentration calculation to the formation energy, where computational errors can lead to an order of magnitude variation in the estimated concentration. Our computational parameters included a cutoff energy of 500 eV and U = 4 eV, whereas a 400 eV cutoff energy and U = 5 eV, all of which could significantly impact the formation energy. Moreover, the absence of experimental values to calibrate the computational results makes it challenging to estimate the magnitude of computational errors, which could be one reason for the substantial discrepancy between the calculated defect concentration and the concentrations claimed by experiments. However, under our computational conditions, it can be qualitatively explained that Cu doping in PPOS can enhance the doping concentration of Cu.

\section{STRUCTURAL AND ELECTRONIC PROPERTIES}

\begin{figure*}[htb]
	\includegraphics[width=0.8\textwidth]{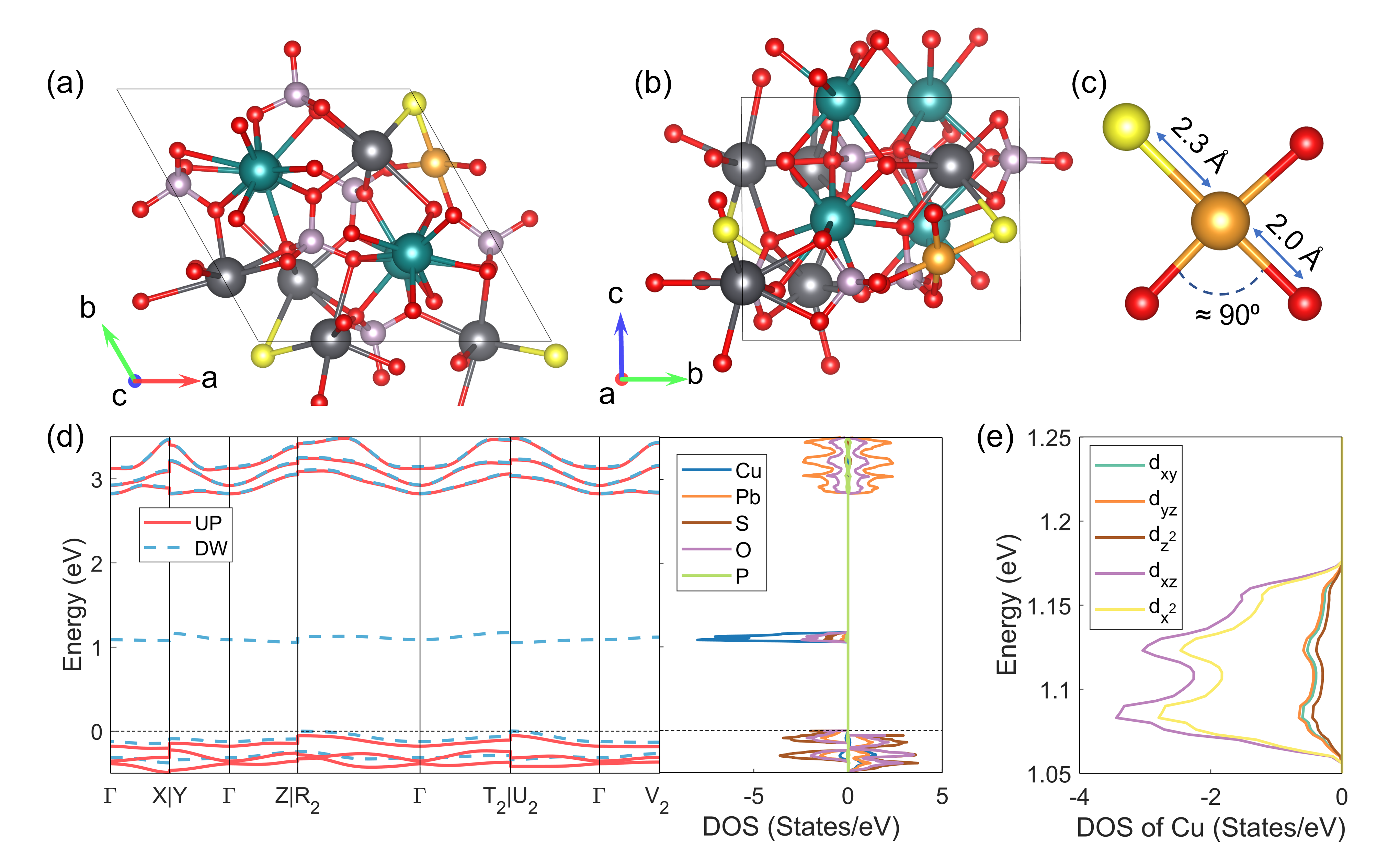}
	\caption{\label{fig4} (a,b) Top and side view of the most likely structure $\mathrm{Pb}_{9}\mathrm{Cu}(\mathrm{PO}_{4})_{6}\mathrm{S}$ (PCPOS). (c) The local structure of PCPOS formed by Cu-O-S. (d) The band structure and DOS of PCPOS, the blue dashed line represent the spin down band and the red solid line represent the band is spin up. (e) DOS of Cu in defect level.}
\end{figure*}

	Given the significant difference in defect formation energy between $x=1$ and $x=2$, it can be inferred that the probability of forming structures with $x=2$ during synthesis is exceedingly low. Consequently, in the doping process within PPOS, the most likely structure to be obtained is $\mathrm{Pb}_{9}\mathrm{Cu}(\mathrm{PO}_{4})_{6}\mathrm{S}$ (PCPOS), as depicted in Fig.~\ref{fig4}(a,b). Interestingly, we have observed a propensity for Cu-O-S to form an approximately planar quadrilateral structure, as shown in Fig.\ref{fig4}(c), which bears resemblance to the Cu-O planes found in high-temperature cuprate superconductors\cite{logvenov2009hightemperature}. The bond angles within this structure are approximately $90^\circ$, with the Cu-O bond length measured at approximately 2.0 $\mathring{\mathrm{A}}$ and the Cu-S bond length at about 2.3 $\mathring{\mathrm{A}}$, which is notably longer than that of the Cu-O bond. Analyze the Bader charge, Cu is in $+1$ oxidation state, S in $-1$ oxidation state, in line with prior research\cite{toriyama2024defect}. In other concentration of Cu doping, the most stable configurations consistently exhibit the formation of such local Cu-O-S structures. 
	
	Nevertheless, the calculated band structures and electronic density of states (DOS) are shown in Fig.~\ref{fig4}(d,e). PCPOS is a semiconductor with 1.1 eV band gap, with a ferromagnetic ground state. The orbital mainly attributed by $d_{xz}$ and $d_{z^2}$. From the band structure, it can be observed that the defect energy levels correspond to downward-spin bands, primarily contributed by Cu, O, and S, corresponding to the planar quadrilateral structure composed of these three elements.
	
	In addition, we have concurrently computed the electronic properties of the most stable configurations for Pb(2) site substitution with different concentrations of $x$ within a single channel, which were all found to be semiconductors. This conclusion is consistent with the results obtained when parent compound is PPO and Cu substitutes for the Pb(2) sites\cite{yang2023initio, sun2023metallization,liu2023symmetry}. Furthermore, we have observed the trends in the changes of band structures at various concentrations, all exhibiting ferromagnetism, with the defect levels primarily contributed by Cu. The results presented herein elucidate that the critical element dictating the electronic characteristics of the material post-synthesis is Cu, while the impact of S on the electronic properties is relatively minor.
	
\section{Discussion}
	
	Incorporating Cu into PPOS is influenced by the presence of S, which can qualitatively address the issue of low Cu doping concentration. However, it enhances the propensity of Cu to substitute Pb(2) sites, resulting in a semiconductor energy band structure. Previous research has discussed that only when Cu replaces Pb(1) sites can a flat band across the Fermi level emerge, which is considered one of the possible characteristics of superconductivity. Therefore, to obtain an appropriate structure for superconductivity during the synthesis process, we should consider how to dope Cu into Pb(1) sites as much as possible, and thus PPOS should not be used as the parent structure for doping.
	
	In recent experiments, the synthesis procedure was adjusted\cite{wang2024observation}, leading to a significant improvement in the sample's performance. The samples were first synthesized through a hydrothermal-calcination method, resulting in a gray, insulating powder. The samples were then ball-milled to nano-scaled particles, soaked in an S solution, and finally subjected to a high-pressure reaction vessel, causing the powder to reorient. 

	Because the defect formation energies of Cu doping in Pb(1) or Pb(2) sites is similar\cite{yang2023initio, sun2023metallization,liu2023symmetry}, we speculate that the purpose of first synthesizing $\mathrm{Pb}_{10-x}\mathrm{Cu}_{x}(\mathrm{PO}_{4})_{6}\mathrm{O}$ is to obtain a structure with Cu partially replacing Pb(1) sites. The magnetic and electric properties emerged only after the powder was separately doped with S, indicating a significant impact of S on electronic properties. The powdering process is speculated to be necessary due to the difficulty and uneven distribution of Cu doping in the sample, as observed in early research\cite{puphal2023single}, where different transparency areas are present. It also suggests the possible presence of a high concentration of S in the final sample. Interestingly, oriented alignment was observed during the compaction process, indicating interactions between different units. However, when S is at the center of the Pb(2) site, it is difficult to generate strong interactions due to the larger distance between the unit and other units, leading to a tendency for disorderly distributed defects. Therefore, the occurrence of oriented alignment suggests that S likely enters PO$_4$ group, utilizing the interactions between S and Cu to attract suitable structures to form pathways.
	
	In summary, we speculate that an appropriate structure for superconductivity may involve Cu occupying Pb(1) sites and S replacing PO$_4$ group, with a possible chemical formula of $\mathrm{Pb}_{10-x}\mathrm{Cu}_{x}(\mathrm{PS}_{y}\mathrm{O}_{4-y})_{6}\mathrm{S}$. Furthermore, since S has little impact on the electronic properties in $\mathrm{Pb}_{10-x}\mathrm{Cu}_{x}(\mathrm{PO}_{4})_{6}\mathrm{O}$, its primary role during the synthesis process is to influence the distribution of Cu. Due to the low concentration of Cu doping, the presence of S allows for the aggregation of suitable Cu-substituted structures, which in turn form a pathway. This phenomenon is manifested as an oriented alignment in the experimental observations.

\section{CONCLUSION}
	
	Our findings elucidate that the transition from PPO to PPOS as the parent material enables stable existence. The emergence of S atoms increases the preference for Cu to substitute at the Pb(2) sites. Additionally, the configuration of Cu between different units has a minimal impact on the defect formation energy, implying that the defects may be distributed in a disordered manner. Based on the formation energy of defects, we predict the defect concentration to be approximately 0.26\%. Qualitatively, the introduction of S can enhance the doping concentration of Cu. Subsequently, we calculated the electronic properties of $\mathrm{Pb}_{10-x}\mathrm{Cu}_{x}(\mathrm{PO}_{4})_{6}\mathrm{O}$ and found that upon doping with Cu, when Cu occupies the Pb(2) site, the material remains a semiconductor, consistent with the electronic properties of Cu occupying the Pb(2) site in PPO\cite{yang2023initio, sun2023metallization,liu2023symmetry}. This indicates that S has a minimal impact on the electronic properties. Finally, in light of our computational results and the experimental synthesis process, we propose that an appropriate structure for superconductivity is likely to be $\mathrm{Pb}_{10-x}\mathrm{Cu}_{x}(\mathrm{PS}_{y}\mathrm{O}_{4-y})_{6}\mathrm{S}$. The role of S in this context is to aggregate the low-concentration Cu defects and form a pathway during the high-pressure reaction vessel process. However, a more systematic and in-depth discussion is required regarding the stoichiometry within $\mathrm{Pb}_{10-x}\mathrm{Cu}_{x}(\mathrm{PS}_{y}\mathrm{O}_{4-y})_{6}\mathrm{S}$, as well as the stability and electronic properties of the corresponding structures.
	
\begin{acknowledgments}
	This work is financially supported by National Natural Science Foundation of China (Grant No. 12074126), Guangdong Basic and Applied Basic Research Foundation (No. 2023A1515012289). The computer times at the High Performance Computational center at South China University of Technology are gratefully acknowledged.
\end{acknowledgments}

	\bibliography{arXiv_v1}

\begin{thebibliography}{30}%
\makeatletter
\providecommand \@ifxundefined [1]{%
 \@ifx{#1\undefined}
}%
\providecommand \@ifnum [1]{%
 \ifnum #1\expandafter \@firstoftwo
 \else \expandafter \@secondoftwo
 \fi
}%
\providecommand \@ifx [1]{%
 \ifx #1\expandafter \@firstoftwo
 \else \expandafter \@secondoftwo
 \fi
}%
\providecommand \natexlab [1]{#1}%
\providecommand \enquote  [1]{``#1''}%
\providecommand \bibnamefont  [1]{#1}%
\providecommand \bibfnamefont [1]{#1}%
\providecommand \citenamefont [1]{#1}%
\providecommand \href@noop [0]{\@secondoftwo}%
\providecommand \href [0]{\begingroup \@sanitize@url \@href}%
\providecommand \@href[1]{\@@startlink{#1}\@@href}%
\providecommand \@@href[1]{\endgroup#1\@@endlink}%
\providecommand \@sanitize@url [0]{\catcode `\\12\catcode `\$12\catcode
  `\&12\catcode `\#12\catcode `\^12\catcode `\_12\catcode `\%12\relax}%
\providecommand \@@startlink[1]{}%
\providecommand \@@endlink[0]{}%
\providecommand \url  [0]{\begingroup\@sanitize@url \@url }%
\providecommand \@url [1]{\endgroup\@href {#1}{\urlprefix }}%
\providecommand \urlprefix  [0]{URL }%
\providecommand \Eprint [0]{\href }%
\providecommand \doibase [0]{https://doi.org/}%
\providecommand \selectlanguage [0]{\@gobble}%
\providecommand \bibinfo  [0]{\@secondoftwo}%
\providecommand \bibfield  [0]{\@secondoftwo}%
\providecommand \translation [1]{[#1]}%
\providecommand \BibitemOpen [0]{}%
\providecommand \bibitemStop [0]{}%
\providecommand \bibitemNoStop [0]{.\EOS\space}%
\providecommand \EOS [0]{\spacefactor3000\relax}%
\providecommand \BibitemShut  [1]{\csname bibitem#1\endcsname}%
\let\auto@bib@innerbib\@empty
\bibitem [{\citenamefont {Lee}\ \emph {et~al.}(2023{\natexlab{a}})\citenamefont
  {Lee}, \citenamefont {Kim},\ and\ \citenamefont {Kwon}}]{lee2023first}%
  \BibitemOpen
  \bibfield  {author} {\bibinfo {author} {\bibfnamefont {S.}~\bibnamefont
  {Lee}}, \bibinfo {author} {\bibfnamefont {J.-H.}\ \bibnamefont {Kim}},\ and\
  \bibinfo {author} {\bibfnamefont {Y.-W.}\ \bibnamefont {Kwon}},\ }\href
  {https://doi.org/10.48550/arXiv.2307.12008} {\bibinfo {title} {The first
  room-temperature ambient-pressure superconductor}} (\bibinfo {year}
  {2023}{\natexlab{a}}),\ \Eprint {https://arxiv.org/abs/2307.12008}
  {arXiv:2307.12008 [cond-mat]} \BibitemShut {NoStop}%
\bibitem [{\citenamefont {Lee}\ \emph {et~al.}(2023{\natexlab{b}})\citenamefont
  {Lee}, \citenamefont {Kim}, \citenamefont {Kim}, \citenamefont {Im},
  \citenamefont {An},\ and\ \citenamefont {Auh}}]{lee2023superconductor}%
  \BibitemOpen
  \bibfield  {author} {\bibinfo {author} {\bibfnamefont {S.}~\bibnamefont
  {Lee}}, \bibinfo {author} {\bibfnamefont {J.}~\bibnamefont {Kim}}, \bibinfo
  {author} {\bibfnamefont {H.-T.}\ \bibnamefont {Kim}}, \bibinfo {author}
  {\bibfnamefont {S.}~\bibnamefont {Im}}, \bibinfo {author} {\bibfnamefont
  {S.}~\bibnamefont {An}},\ and\ \bibinfo {author} {\bibfnamefont {K.~H.}\
  \bibnamefont {Auh}},\ }\href {https://doi.org/10.48550/arXiv.2307.12037}
  {\bibinfo {title} {Superconductor
  ${\mathrm{pb}}_{10-x}\mathrm{Cu}_{x}{({\mathrm{PO}}_{4})}_{6}\mathrm{O}$
  showing levitation at room temperature and atmospheric pressure and
  mechanism}} (\bibinfo {year} {2023}{\natexlab{b}}),\ \Eprint
  {https://arxiv.org/abs/2307.12037} {arXiv:2307.12037 [cond-mat]} \BibitemShut
  {NoStop}%
\bibitem [{\citenamefont {Puphal}\ \emph {et~al.}(2023)\citenamefont {Puphal},
  \citenamefont {Akbar}, \citenamefont {Hepting}, \citenamefont {Goering},
  \citenamefont {Isobe}, \citenamefont {Nugroho},\ and\ \citenamefont
  {Keimer}}]{puphal2023single}%
  \BibitemOpen
  \bibfield  {author} {\bibinfo {author} {\bibfnamefont {P.}~\bibnamefont
  {Puphal}}, \bibinfo {author} {\bibfnamefont {M.~Y.~P.}\ \bibnamefont
  {Akbar}}, \bibinfo {author} {\bibfnamefont {M.}~\bibnamefont {Hepting}},
  \bibinfo {author} {\bibfnamefont {E.}~\bibnamefont {Goering}}, \bibinfo
  {author} {\bibfnamefont {M.}~\bibnamefont {Isobe}}, \bibinfo {author}
  {\bibfnamefont {A.~A.}\ \bibnamefont {Nugroho}},\ and\ \bibinfo {author}
  {\bibfnamefont {B.}~\bibnamefont {Keimer}},\ }\bibfield  {title} {\bibinfo
  {title} {Single crystal synthesis, structure, and magnetism of
  {${\mathrm{Pb}}_{10-x}\mathrm{Cu}_{x}{({\mathrm{PO}}_{4})}_{6}\mathrm{O}$}},\
  }\bibfield  {journal} {\bibinfo  {journal} {APL Materials}\ }\textbf
  {\bibinfo {volume} {11}},\ \href {https://doi.org/10.1063/5.0172755}
  {10.1063/5.0172755} (\bibinfo {year} {2023})\BibitemShut {NoStop}%
\bibitem [{\citenamefont {Liu}\ \emph {et~al.}(2023{\natexlab{a}})\citenamefont
  {Liu}, \citenamefont {Meng}, \citenamefont {Wang}, \citenamefont {Chen},
  \citenamefont {Duan}, \citenamefont {Zhou}, \citenamefont {Yan},
  \citenamefont {Qin},\ and\ \citenamefont {Liu}}]{liu2023semiconducting}%
  \BibitemOpen
  \bibfield  {author} {\bibinfo {author} {\bibfnamefont {L.}~\bibnamefont
  {Liu}}, \bibinfo {author} {\bibfnamefont {Z.}~\bibnamefont {Meng}}, \bibinfo
  {author} {\bibfnamefont {X.}~\bibnamefont {Wang}}, \bibinfo {author}
  {\bibfnamefont {H.}~\bibnamefont {Chen}}, \bibinfo {author} {\bibfnamefont
  {Z.}~\bibnamefont {Duan}}, \bibinfo {author} {\bibfnamefont {X.}~\bibnamefont
  {Zhou}}, \bibinfo {author} {\bibfnamefont {H.}~\bibnamefont {Yan}}, \bibinfo
  {author} {\bibfnamefont {P.}~\bibnamefont {Qin}},\ and\ \bibinfo {author}
  {\bibfnamefont {Z.}~\bibnamefont {Liu}},\ }\bibfield  {title} {\bibinfo
  {title} {Semiconducting transport in
  {${\mathrm{Pb}}_{10-x}\mathrm{Cu}_{x}{({\mathrm{PO}}_{4})}_{6}\mathrm{O}$}
  sintered from {$\mathrm{Pb}_2\mathrm{SO}_5$} and {Cu$_3$P}},\ }\href
  {https://doi.org/10.1002/adfm.202308938} {\bibfield  {journal} {\bibinfo
  {journal} {Advanced Functional Materials}\ }\textbf {\bibinfo {volume}
  {33}},\ \bibinfo {pages} {2308938} (\bibinfo {year}
  {2023}{\natexlab{a}})}\BibitemShut {NoStop}%
\bibitem [{\citenamefont {Wang}\ \emph {et~al.}(2023)\citenamefont {Wang},
  \citenamefont {Liu}, \citenamefont {Ge}, \citenamefont {Ji}, \citenamefont
  {Ji}, \citenamefont {Liu}, \citenamefont {Ai}, \citenamefont {Ma},
  \citenamefont {Qi},\ and\ \citenamefont {Wang}}]{wang2023ferromagnetic}%
  \BibitemOpen
  \bibfield  {author} {\bibinfo {author} {\bibfnamefont {P.}~\bibnamefont
  {Wang}}, \bibinfo {author} {\bibfnamefont {X.}~\bibnamefont {Liu}}, \bibinfo
  {author} {\bibfnamefont {J.}~\bibnamefont {Ge}}, \bibinfo {author}
  {\bibfnamefont {C.}~\bibnamefont {Ji}}, \bibinfo {author} {\bibfnamefont
  {H.}~\bibnamefont {Ji}}, \bibinfo {author} {\bibfnamefont {Y.}~\bibnamefont
  {Liu}}, \bibinfo {author} {\bibfnamefont {Y.}~\bibnamefont {Ai}}, \bibinfo
  {author} {\bibfnamefont {G.}~\bibnamefont {Ma}}, \bibinfo {author}
  {\bibfnamefont {S.}~\bibnamefont {Qi}},\ and\ \bibinfo {author}
  {\bibfnamefont {J.}~\bibnamefont {Wang}},\ }\bibfield  {title} {\bibinfo
  {title} {Ferromagnetic and insulating behavior in both half magnetic
  levitation and non-levitation {LK}-99 like samples},\ }\href
  {https://doi.org/10.1007/s44214-023-00035-z} {\bibfield  {journal} {\bibinfo
  {journal} {Quantum Front}\ }\textbf {\bibinfo {volume} {2}},\ \bibinfo
  {pages} {10} (\bibinfo {year} {2023})}\BibitemShut {NoStop}%
\bibitem [{\citenamefont {Kumar}\ \emph {et~al.}(2023)\citenamefont {Kumar},
  \citenamefont {Karn}, \citenamefont {Kumar},\ and\ \citenamefont
  {Awana}}]{kumar2023absence}%
  \BibitemOpen
  \bibfield  {author} {\bibinfo {author} {\bibfnamefont {K.}~\bibnamefont
  {Kumar}}, \bibinfo {author} {\bibfnamefont {N.~K.}\ \bibnamefont {Karn}},
  \bibinfo {author} {\bibfnamefont {Y.}~\bibnamefont {Kumar}},\ and\ \bibinfo
  {author} {\bibfnamefont {V.~P.~S.}\ \bibnamefont {Awana}},\ }\bibfield
  {title} {\bibinfo {title} {Absence of superconductivity in {LK-99} at ambient
  conditions},\ }\href {https://doi.org/10.1021/acsomega.3c06096} {\bibfield
  {journal} {\bibinfo  {journal} {ACS Omega}\ }\textbf {\bibinfo {volume}
  {8}},\ \bibinfo {pages} {41737} (\bibinfo {year} {2023})}\BibitemShut
  {NoStop}%
\bibitem [{\citenamefont {Zhu}\ \emph {et~al.}(2023)\citenamefont {Zhu},
  \citenamefont {Wu}, \citenamefont {Li},\ and\ \citenamefont
  {Luo}}]{zhu2023firstorder}%
  \BibitemOpen
  \bibfield  {author} {\bibinfo {author} {\bibfnamefont {S.}~\bibnamefont
  {Zhu}}, \bibinfo {author} {\bibfnamefont {W.}~\bibnamefont {Wu}}, \bibinfo
  {author} {\bibfnamefont {Z.}~\bibnamefont {Li}},\ and\ \bibinfo {author}
  {\bibfnamefont {J.}~\bibnamefont {Luo}},\ }\bibfield  {title} {\bibinfo
  {title} {First-order transition in {LK-99} containing {Cu$_2$S}},\ }\href
  {https://doi.org/10.1016/j.matt.2023.11.001} {\bibfield  {journal} {\bibinfo
  {journal} {Matter}\ }\textbf {\bibinfo {volume} {6}},\ \bibinfo {pages}
  {4401} (\bibinfo {year} {2023})}\BibitemShut {NoStop}%
\bibitem [{\citenamefont {Timokhin}\ \emph {et~al.}(2023)\citenamefont
  {Timokhin}, \citenamefont {Chen}, \citenamefont {Wang}, \citenamefont
  {Yang},\ and\ \citenamefont {Mishchenko}}]{timokhin2023synthesis}%
  \BibitemOpen
  \bibfield  {author} {\bibinfo {author} {\bibfnamefont {I.}~\bibnamefont
  {Timokhin}}, \bibinfo {author} {\bibfnamefont {C.}~\bibnamefont {Chen}},
  \bibinfo {author} {\bibfnamefont {Z.}~\bibnamefont {Wang}}, \bibinfo {author}
  {\bibfnamefont {Q.}~\bibnamefont {Yang}},\ and\ \bibinfo {author}
  {\bibfnamefont {A.}~\bibnamefont {Mishchenko}},\ }\href
  {https://doi.org/10.48550/arXiv.2308.03823} {\bibinfo {title} {Synthesis and
  characterisation of {LK-99}}} (\bibinfo {year} {2023}),\ \Eprint
  {https://arxiv.org/abs/2308.03823} {arXiv:2308.03823 [cond-mat]} \BibitemShut
  {NoStop}%
\bibitem [{\citenamefont {Zhang}\ \emph {et~al.}(2024)\citenamefont {Zhang},
  \citenamefont {Liu}, \citenamefont {Zhu},\ and\ \citenamefont
  {Wen}}]{zhang2024ferromagnetism}%
  \BibitemOpen
  \bibfield  {author} {\bibinfo {author} {\bibfnamefont {Y.}~\bibnamefont
  {Zhang}}, \bibinfo {author} {\bibfnamefont {C.}~\bibnamefont {Liu}}, \bibinfo
  {author} {\bibfnamefont {X.}~\bibnamefont {Zhu}},\ and\ \bibinfo {author}
  {\bibfnamefont {H.-H.}\ \bibnamefont {Wen}},\ }\bibfield  {title} {\bibinfo
  {title} {Ferromagnetism and insulating behavior with a logarithmic
  temperature dependence of resistivity in
  {${\mathrm{Pb}}_{10-x}\mathrm{Cu}_{x}{({\mathrm{PO}}_{4})}_{6}\mathrm{O}$}},\
  }\href {https://doi.org/10.1007/s11433-023-2209-7} {\bibfield  {journal}
  {\bibinfo  {journal} {Sci. China Phys. Mech. Astron.}\ }\textbf {\bibinfo
  {volume} {67}},\ \bibinfo {pages} {217413} (\bibinfo {year} {2024})},\
  \Eprint {https://arxiv.org/abs/2308.05786} {arXiv:2308.05786 [cond-mat]}
  \BibitemShut {NoStop}%
\bibitem [{\citenamefont {Thakur}\ \emph {et~al.}(2023)\citenamefont {Thakur},
  \citenamefont {Schulze},\ and\ \citenamefont {Ruck}}]{thakur2023synthesis}%
  \BibitemOpen
  \bibfield  {author} {\bibinfo {author} {\bibfnamefont {G.~S.}\ \bibnamefont
  {Thakur}}, \bibinfo {author} {\bibfnamefont {M.}~\bibnamefont {Schulze}},\
  and\ \bibinfo {author} {\bibfnamefont {M.}~\bibnamefont {Ruck}},\ }\bibfield
  {title} {\bibinfo {title} {On the synthesis methodologies to prepare
  {${\mathrm{Pb}}_{9}\mathrm{Cu}{({\mathrm{PO}}_{4})}_{6}\mathrm{O}$}: Phase,
  composition, magnetic analysis and absence of superconductivity},\ }\href
  {https://doi.org/10.1088/1361-6668/ad1250} {\bibfield  {journal} {\bibinfo
  {journal} {Supercond. Sci. Technol.}\ }\textbf {\bibinfo {volume} {37}},\
  \bibinfo {pages} {015013} (\bibinfo {year} {2023})}\BibitemShut {NoStop}%
\bibitem [{\citenamefont {Lei}\ \emph {et~al.}(2024)\citenamefont {Lei},
  \citenamefont {Lin}, \citenamefont {Chen}, \citenamefont {Chou},
  \citenamefont {Lin}, \citenamefont {Chen}, \citenamefont {Sung},\ and\
  \citenamefont {Wang}}]{lei2024characteristics}%
  \BibitemOpen
  \bibfield  {author} {\bibinfo {author} {\bibfnamefont {Z.}~\bibnamefont
  {Lei}}, \bibinfo {author} {\bibfnamefont {C.-W.}\ \bibnamefont {Lin}},
  \bibinfo {author} {\bibfnamefont {I.-N.}\ \bibnamefont {Chen}}, \bibinfo
  {author} {\bibfnamefont {C.-T.}\ \bibnamefont {Chou}}, \bibinfo {author}
  {\bibfnamefont {Y.-L.}\ \bibnamefont {Lin}}, \bibinfo {author} {\bibfnamefont
  {J.-H.}\ \bibnamefont {Chen}}, \bibinfo {author} {\bibfnamefont {H.-H.}\
  \bibnamefont {Sung}},\ and\ \bibinfo {author} {\bibfnamefont {L.-M.}\
  \bibnamefont {Wang}},\ }\bibfield  {title} {\bibinfo {title} {The
  characteristics of cu-doped lead apatite (lk-99) synthesized with the removal
  of cu2s using ammonia solution: A diamagnetic semiconductor},\ }\href
  {https://doi.org/10.1063/5.0183271} {\bibfield  {journal} {\bibinfo
  {journal} {APL Materials}\ }\textbf {\bibinfo {volume} {12}},\ \bibinfo
  {pages} {021104} (\bibinfo {year} {2024})}\BibitemShut {NoStop}%
\bibitem [{\citenamefont {Toriyama}\ \emph {et~al.}(2024)\citenamefont
  {Toriyama}, \citenamefont {Lee}, \citenamefont {Snyder},\ and\ \citenamefont
  {Gorai}}]{toriyama2024defect}%
  \BibitemOpen
  \bibfield  {author} {\bibinfo {author} {\bibfnamefont {M.~Y.}\ \bibnamefont
  {Toriyama}}, \bibinfo {author} {\bibfnamefont {C.-W.}\ \bibnamefont {Lee}},
  \bibinfo {author} {\bibfnamefont {G.~J.}\ \bibnamefont {Snyder}},\ and\
  \bibinfo {author} {\bibfnamefont {P.}~\bibnamefont {Gorai}},\ }\bibfield
  {title} {\bibinfo {title} {Defect chemistry and doping of lead phosphate oxo
  apatite {${\mathrm{Pb}}_{10}{({\mathrm{PO}}_{4})}_{6}\mathrm{O}$}},\ }\href
  {https://doi.org/10.1021/acsenergylett.3c02544} {\bibfield  {journal}
  {\bibinfo  {journal} {ACS Energy Lett.}\ ,\ \bibinfo {pages} {428}} (\bibinfo
  {year} {2024})}\BibitemShut {NoStop}%
\bibitem [{\citenamefont {Jain}(2023)}]{jain2023superionica}%
  \BibitemOpen
  \bibfield  {author} {\bibinfo {author} {\bibfnamefont {P.~K.}\ \bibnamefont
  {Jain}},\ }\bibfield  {title} {\bibinfo {title} {Superionic phase transition
  of copper(i) sulfide and its implication for purported superconductivity of
  {LK}-99},\ }\href {https://doi.org/10.1021/acs.jpcc.3c05684} {\bibfield
  {journal} {\bibinfo  {journal} {J. Phys. Chem. C}\ }\textbf {\bibinfo
  {volume} {127}},\ \bibinfo {pages} {18253} (\bibinfo {year} {2023})},\
  \Eprint {https://arxiv.org/abs/2308.05222} {arXiv:2308.05222 [cond-mat]}
  \BibitemShut {NoStop}%
\bibitem [{\citenamefont {Liu}\ \emph {et~al.}(2023{\natexlab{b}})\citenamefont
  {Liu}, \citenamefont {Cheng}, \citenamefont {Zhang}, \citenamefont {Xu},
  \citenamefont {Li}, \citenamefont {Shi}, \citenamefont {Yuan}, \citenamefont
  {Xu}, \citenamefont {Zhou}, \citenamefont {Zhu}, \citenamefont {Sun},
  \citenamefont {Wu}, \citenamefont {Luo}, \citenamefont {Jin},\ and\
  \citenamefont {Li}}]{liu2023phases}%
  \BibitemOpen
  \bibfield  {author} {\bibinfo {author} {\bibfnamefont {C.}~\bibnamefont
  {Liu}}, \bibinfo {author} {\bibfnamefont {W.}~\bibnamefont {Cheng}}, \bibinfo
  {author} {\bibfnamefont {X.}~\bibnamefont {Zhang}}, \bibinfo {author}
  {\bibfnamefont {J.}~\bibnamefont {Xu}}, \bibinfo {author} {\bibfnamefont
  {J.}~\bibnamefont {Li}}, \bibinfo {author} {\bibfnamefont {Q.}~\bibnamefont
  {Shi}}, \bibinfo {author} {\bibfnamefont {C.}~\bibnamefont {Yuan}}, \bibinfo
  {author} {\bibfnamefont {L.}~\bibnamefont {Xu}}, \bibinfo {author}
  {\bibfnamefont {H.}~\bibnamefont {Zhou}}, \bibinfo {author} {\bibfnamefont
  {S.}~\bibnamefont {Zhu}}, \bibinfo {author} {\bibfnamefont {J.}~\bibnamefont
  {Sun}}, \bibinfo {author} {\bibfnamefont {W.}~\bibnamefont {Wu}}, \bibinfo
  {author} {\bibfnamefont {J.}~\bibnamefont {Luo}}, \bibinfo {author}
  {\bibfnamefont {K.}~\bibnamefont {Jin}},\ and\ \bibinfo {author}
  {\bibfnamefont {Y.}~\bibnamefont {Li}},\ }\bibfield  {title} {\bibinfo
  {title} {Phases and magnetism at microscale in compounds containing nominal
  {${\mathrm{Pb}}_{10-x}\mathrm{Cu}_{x}{({\mathrm{PO}}_{4})}_{6}\mathrm{O}$}},\
  }\href {https://doi.org/10.1103/PhysRevMaterials.7.084804} {\bibfield
  {journal} {\bibinfo  {journal} {Phys. Rev. Mater.}\ }\textbf {\bibinfo
  {volume} {7}},\ \bibinfo {pages} {084804} (\bibinfo {year}
  {2023}{\natexlab{b}})}\BibitemShut {NoStop}%
\bibitem [{\citenamefont {Yang}\ \emph {et~al.}(2023)\citenamefont {Yang},
  \citenamefont {Liu},\ and\ \citenamefont {Zhong}}]{yang2023initio}%
  \BibitemOpen
  \bibfield  {author} {\bibinfo {author} {\bibfnamefont {S.}~\bibnamefont
  {Yang}}, \bibinfo {author} {\bibfnamefont {G.}~\bibnamefont {Liu}},\ and\
  \bibinfo {author} {\bibfnamefont {Y.}~\bibnamefont {Zhong}},\ }\bibfield
  {title} {\bibinfo {title} {Ab initio investigations on the electronic
  properties and stability of {Cu}-substituted lead apatite ({LK-99}) family
  with different doping concentrations ({$x=0, 1, 2$})},\ }\href
  {https://doi.org/10.1016/j.mtcomm.2023.107379} {\bibfield  {journal}
  {\bibinfo  {journal} {Materials Today Communications}\ }\textbf {\bibinfo
  {volume} {37}},\ \bibinfo {pages} {107379} (\bibinfo {year}
  {2023})}\BibitemShut {NoStop}%
\bibitem [{\citenamefont {Sun}\ \emph {et~al.}(2023)\citenamefont {Sun},
  \citenamefont {Ho},\ and\ \citenamefont {Antropov}}]{sun2023metallization}%
  \BibitemOpen
  \bibfield  {author} {\bibinfo {author} {\bibfnamefont {Y.}~\bibnamefont
  {Sun}}, \bibinfo {author} {\bibfnamefont {K.-M.}\ \bibnamefont {Ho}},\ and\
  \bibinfo {author} {\bibfnamefont {V.}~\bibnamefont {Antropov}},\ }\bibfield
  {title} {\bibinfo {title} {Metallization and spin fluctuations in {Cu}-doped
  lead apatite},\ }\href {https://doi.org/10.1103/PhysRevMaterials.7.114804}
  {\bibfield  {journal} {\bibinfo  {journal} {Phys. Rev. Mater.}\ }\textbf
  {\bibinfo {volume} {7}},\ \bibinfo {pages} {114804} (\bibinfo {year}
  {2023})}\BibitemShut {NoStop}%
\bibitem [{\citenamefont {Korotin}\ \emph {et~al.}(2023)\citenamefont
  {Korotin}, \citenamefont {Novoselov}, \citenamefont {Shorikov}, \citenamefont
  {Anisimov},\ and\ \citenamefont {Oganov}}]{korotin2023electronic}%
  \BibitemOpen
  \bibfield  {author} {\bibinfo {author} {\bibfnamefont {D.~M.}\ \bibnamefont
  {Korotin}}, \bibinfo {author} {\bibfnamefont {D.~Y.}\ \bibnamefont
  {Novoselov}}, \bibinfo {author} {\bibfnamefont {A.~O.}\ \bibnamefont
  {Shorikov}}, \bibinfo {author} {\bibfnamefont {V.~I.}\ \bibnamefont
  {Anisimov}},\ and\ \bibinfo {author} {\bibfnamefont {A.~R.}\ \bibnamefont
  {Oganov}},\ }\bibfield  {title} {\bibinfo {title} {Electronic correlations in
  the ultranarrow energy band compound
  {${\mathrm{Pb}}_{9}\mathrm{Cu}{({\mathrm{PO}}_{4})}_{6}\mathrm{O}$}: A
  {$\mathrm{DFT}+\mathrm{DMFT}$} study},\ }\href
  {https://doi.org/10.1103/PhysRevB.108.L241111} {\bibfield  {journal}
  {\bibinfo  {journal} {Phys. Rev. B}\ }\textbf {\bibinfo {volume} {108}},\
  \bibinfo {pages} {L241111} (\bibinfo {year} {2023})}\BibitemShut {NoStop}%
\bibitem [{\citenamefont {Si}\ and\ \citenamefont
  {Held}(2023)}]{si2023electronic}%
  \BibitemOpen
  \bibfield  {author} {\bibinfo {author} {\bibfnamefont {L.}~\bibnamefont
  {Si}}\ and\ \bibinfo {author} {\bibfnamefont {K.}~\bibnamefont {Held}},\
  }\bibfield  {title} {\bibinfo {title} {Electronic structure of the putative
  room-temperature superconductor
  {${\mathrm{Pb}}_{9}\mathrm{Cu}{({\mathrm{PO}}_{4})}_{6}\mathrm{O}$}},\ }\href
  {https://doi.org/10.1103/PhysRevB.108.L121110} {\bibfield  {journal}
  {\bibinfo  {journal} {Phys. Rev. B}\ }\textbf {\bibinfo {volume} {108}},\
  \bibinfo {pages} {L121110} (\bibinfo {year} {2023})}\BibitemShut {NoStop}%
\bibitem [{\citenamefont {Liu}\ \emph {et~al.}(2023{\natexlab{c}})\citenamefont
  {Liu}, \citenamefont {Yu}, \citenamefont {Li}, \citenamefont {Wang},
  \citenamefont {Lai}, \citenamefont {Sun}, \citenamefont {Chen},\ and\
  \citenamefont {Liu}}]{liu2023symmetry}%
  \BibitemOpen
  \bibfield  {author} {\bibinfo {author} {\bibfnamefont {J.}~\bibnamefont
  {Liu}}, \bibinfo {author} {\bibfnamefont {T.}~\bibnamefont {Yu}}, \bibinfo
  {author} {\bibfnamefont {J.}~\bibnamefont {Li}}, \bibinfo {author}
  {\bibfnamefont {J.}~\bibnamefont {Wang}}, \bibinfo {author} {\bibfnamefont
  {J.}~\bibnamefont {Lai}}, \bibinfo {author} {\bibfnamefont {Y.}~\bibnamefont
  {Sun}}, \bibinfo {author} {\bibfnamefont {X.-Q.}\ \bibnamefont {Chen}},\ and\
  \bibinfo {author} {\bibfnamefont {P.}~\bibnamefont {Liu}},\ }\bibfield
  {title} {\bibinfo {title} {Symmetry breaking induced insulating electronic
  state in
  {${\mathrm{Pb}}_{9}\mathrm{Cu}{({\mathrm{PO}}_{4})}_{6}\mathrm{O}$}},\ }\href
  {https://doi.org/10.1103/PhysRevB.108.L161101} {\bibfield  {journal}
  {\bibinfo  {journal} {Phys. Rev. B}\ }\textbf {\bibinfo {volume} {108}},\
  \bibinfo {pages} {L161101} (\bibinfo {year}
  {2023}{\natexlab{c}})}\BibitemShut {NoStop}%
\bibitem [{\citenamefont {Wang}\ \emph
  {et~al.}(2024{\natexlab{a}})\citenamefont {Wang}, \citenamefont {Yao},
  \citenamefont {Shi}, \citenamefont {Zhao}, \citenamefont {Wu}, \citenamefont
  {Wu}, \citenamefont {Geng}, \citenamefont {Ye},\ and\ \citenamefont
  {Chen}}]{wang2024possible}%
  \BibitemOpen
  \bibfield  {author} {\bibinfo {author} {\bibfnamefont {H.}~\bibnamefont
  {Wang}}, \bibinfo {author} {\bibfnamefont {Y.}~\bibnamefont {Yao}}, \bibinfo
  {author} {\bibfnamefont {K.}~\bibnamefont {Shi}}, \bibinfo {author}
  {\bibfnamefont {Y.}~\bibnamefont {Zhao}}, \bibinfo {author} {\bibfnamefont
  {H.}~\bibnamefont {Wu}}, \bibinfo {author} {\bibfnamefont {Z.}~\bibnamefont
  {Wu}}, \bibinfo {author} {\bibfnamefont {Z.}~\bibnamefont {Geng}}, \bibinfo
  {author} {\bibfnamefont {S.}~\bibnamefont {Ye}},\ and\ \bibinfo {author}
  {\bibfnamefont {N.}~\bibnamefont {Chen}},\ }\href@noop {} {\bibinfo {title}
  {Possible {Meissner} effect near room temperature in copper-substituted lead
  apatite}} (\bibinfo {year} {2024}{\natexlab{a}}),\ \Eprint
  {https://arxiv.org/abs/2401.00999} {arXiv:2401.00999 [cond-mat]} \BibitemShut
  {NoStop}%
\bibitem [{\citenamefont {Wang}\ \emph
  {et~al.}(2024{\natexlab{b}})\citenamefont {Wang}, \citenamefont {Wu},
  \citenamefont {Chen}, \citenamefont {Qiao}, \citenamefont {Wang},
  \citenamefont {Wu}, \citenamefont {Geng}, \citenamefont {Xue}, \citenamefont
  {Chang}, \citenamefont {Ye},\ and\ \citenamefont
  {Yao}}]{wang2024observation}%
  \BibitemOpen
  \bibfield  {author} {\bibinfo {author} {\bibfnamefont {H.}~\bibnamefont
  {Wang}}, \bibinfo {author} {\bibfnamefont {H.}~\bibnamefont {Wu}}, \bibinfo
  {author} {\bibfnamefont {N.}~\bibnamefont {Chen}}, \bibinfo {author}
  {\bibfnamefont {X.}~\bibnamefont {Qiao}}, \bibinfo {author} {\bibfnamefont
  {L.}~\bibnamefont {Wang}}, \bibinfo {author} {\bibfnamefont {Z.}~\bibnamefont
  {Wu}}, \bibinfo {author} {\bibfnamefont {Z.}~\bibnamefont {Geng}}, \bibinfo
  {author} {\bibfnamefont {W.}~\bibnamefont {Xue}}, \bibinfo {author}
  {\bibfnamefont {H.}~\bibnamefont {Chang}}, \bibinfo {author} {\bibfnamefont
  {S.}~\bibnamefont {Ye}},\ and\ \bibinfo {author} {\bibfnamefont
  {Y.}~\bibnamefont {Yao}},\ }\href {https://doi.org/10.48550/arXiv.2403.11126}
  {\bibinfo {title} {Observation of diamagnetic strange-metal phase in
  sulfur-copper codoped lead apatite}} (\bibinfo {year} {2024}{\natexlab{b}}),\
  \Eprint {https://arxiv.org/abs/2403.11126} {arXiv:2403.11126 [cond-mat]}
  \BibitemShut {NoStop}%
\bibitem [{\citenamefont {Tao}\ \emph {et~al.}(2024)\citenamefont {Tao},
  \citenamefont {Chen}, \citenamefont {Yang}, \citenamefont {Gao},
  \citenamefont {Xue},\ and\ \citenamefont {Jia}}]{tao2024occupied}%
  \BibitemOpen
  \bibfield  {author} {\bibinfo {author} {\bibfnamefont {K.}~\bibnamefont
  {Tao}}, \bibinfo {author} {\bibfnamefont {R.}~\bibnamefont {Chen}}, \bibinfo
  {author} {\bibfnamefont {L.}~\bibnamefont {Yang}}, \bibinfo {author}
  {\bibfnamefont {J.}~\bibnamefont {Gao}}, \bibinfo {author} {\bibfnamefont
  {D.}~\bibnamefont {Xue}},\ and\ \bibinfo {author} {\bibfnamefont
  {C.}~\bibnamefont {Jia}},\ }\bibfield  {title} {\bibinfo {title} {The 1/4
  occupied {O} atoms induced ultra-flatband and the one-dimensional channels in
  the {${\mathrm{Pb}}_{10-x}\mathrm{Cu}_{x}{({\mathrm{PO}}_{4})}_{6}\mathrm{O}
  \ (x = 0, 0.5)$} crystal},\ }\href {https://doi.org/10.1063/5.0188943}
  {\bibfield  {journal} {\bibinfo  {journal} {APL Materials}\ }\textbf
  {\bibinfo {volume} {12}},\ \bibinfo {pages} {021120} (\bibinfo {year}
  {2024})}\BibitemShut {NoStop}%
\bibitem [{\citenamefont {He}\ \emph {et~al.}(2020)\citenamefont {He},
  \citenamefont {Qiu}, \citenamefont {Yu}, \citenamefont {Liao}, \citenamefont
  {Zhao},\ and\ \citenamefont {Yang}}]{he2020atom}%
  \BibitemOpen
  \bibfield  {author} {\bibinfo {author} {\bibfnamefont {C.-C.}\ \bibnamefont
  {He}}, \bibinfo {author} {\bibfnamefont {S.-B.}\ \bibnamefont {Qiu}},
  \bibinfo {author} {\bibfnamefont {J.-S.}\ \bibnamefont {Yu}}, \bibinfo
  {author} {\bibfnamefont {J.-H.}\ \bibnamefont {Liao}}, \bibinfo {author}
  {\bibfnamefont {Y.-J.}\ \bibnamefont {Zhao}},\ and\ \bibinfo {author}
  {\bibfnamefont {X.-B.}\ \bibnamefont {Yang}},\ }\bibfield  {title} {\bibinfo
  {title} {Atom classification model for total energy evaluation of
  two-dimensional multicomponent materials},\ }\href
  {https://doi.org/10.1021/acs.jpca.0c02431} {\bibfield  {journal} {\bibinfo
  {journal} {J. Phys. Chem. A}\ }\textbf {\bibinfo {volume} {124}},\ \bibinfo
  {pages} {4506} (\bibinfo {year} {2020})}\BibitemShut {NoStop}%
\bibitem [{\citenamefont {He}\ \emph {et~al.}(2021)\citenamefont {He},
  \citenamefont {Liao}, \citenamefont {Qiu}, \citenamefont {Zhao},\ and\
  \citenamefont {Yang}}]{he2021biased}%
  \BibitemOpen
  \bibfield  {author} {\bibinfo {author} {\bibfnamefont {C.-C.}\ \bibnamefont
  {He}}, \bibinfo {author} {\bibfnamefont {J.-H.}\ \bibnamefont {Liao}},
  \bibinfo {author} {\bibfnamefont {S.-B.}\ \bibnamefont {Qiu}}, \bibinfo
  {author} {\bibfnamefont {Y.-J.}\ \bibnamefont {Zhao}},\ and\ \bibinfo
  {author} {\bibfnamefont {X.-B.}\ \bibnamefont {Yang}},\ }\bibfield  {title}
  {\bibinfo {title} {Biased screening for multi-component materials with
  structures of alloy generation and recognition ({SAGAR})},\ }\href
  {https://doi.org/10.1016/j.commatsci.2021.110386} {\bibfield  {journal}
  {\bibinfo  {journal} {Computational Materials Science}\ }\textbf {\bibinfo
  {volume} {193}},\ \bibinfo {pages} {110386} (\bibinfo {year}
  {2021})}\BibitemShut {NoStop}%
\bibitem [{\citenamefont {Kresse}\ and\ \citenamefont
  {Furthm{\"u}ller}(1996{\natexlab{a}})}]{kresse1996efficiency}%
  \BibitemOpen
  \bibfield  {author} {\bibinfo {author} {\bibfnamefont {G.}~\bibnamefont
  {Kresse}}\ and\ \bibinfo {author} {\bibfnamefont {J.}~\bibnamefont
  {Furthm{\"u}ller}},\ }\bibfield  {title} {\bibinfo {title} {Efficiency of
  ab-initio total energy calculations for metals and semiconductors using a
  plane-wave basis set},\ }\href {https://doi.org/10.1016/0927-0256(96)00008-0}
  {\bibfield  {journal} {\bibinfo  {journal} {Computational Materials Science}\
  }\textbf {\bibinfo {volume} {6}},\ \bibinfo {pages} {15} (\bibinfo {year}
  {1996}{\natexlab{a}})}\BibitemShut {NoStop}%
\bibitem [{\citenamefont {Kresse}\ and\ \citenamefont
  {Furthm{\"u}ller}(1996{\natexlab{b}})}]{kresse1996efficient}%
  \BibitemOpen
  \bibfield  {author} {\bibinfo {author} {\bibfnamefont {G.}~\bibnamefont
  {Kresse}}\ and\ \bibinfo {author} {\bibfnamefont {J.}~\bibnamefont
  {Furthm{\"u}ller}},\ }\bibfield  {title} {\bibinfo {title} {Efficient
  iterative schemes for ab initio total-energy calculations using a plane-wave
  basis set},\ }\href {https://doi.org/10.1103/PhysRevB.54.11169} {\bibfield
  {journal} {\bibinfo  {journal} {Phys. Rev. B}\ }\textbf {\bibinfo {volume}
  {54}},\ \bibinfo {pages} {11169} (\bibinfo {year}
  {1996}{\natexlab{b}})}\BibitemShut {NoStop}%
\bibitem [{\citenamefont {Perdew}\ \emph {et~al.}(1992)\citenamefont {Perdew},
  \citenamefont {Chevary}, \citenamefont {Vosko}, \citenamefont {Jackson},
  \citenamefont {Pederson}, \citenamefont {Singh},\ and\ \citenamefont
  {Fiolhais}}]{perdew1992atoms}%
  \BibitemOpen
  \bibfield  {author} {\bibinfo {author} {\bibfnamefont {J.~P.}\ \bibnamefont
  {Perdew}}, \bibinfo {author} {\bibfnamefont {J.~A.}\ \bibnamefont {Chevary}},
  \bibinfo {author} {\bibfnamefont {S.~H.}\ \bibnamefont {Vosko}}, \bibinfo
  {author} {\bibfnamefont {K.~A.}\ \bibnamefont {Jackson}}, \bibinfo {author}
  {\bibfnamefont {M.~R.}\ \bibnamefont {Pederson}}, \bibinfo {author}
  {\bibfnamefont {D.~J.}\ \bibnamefont {Singh}},\ and\ \bibinfo {author}
  {\bibfnamefont {C.}~\bibnamefont {Fiolhais}},\ }\bibfield  {title} {\bibinfo
  {title} {Atoms, molecules, solids, and surfaces: Applications of the
  generalized gradient approximation for exchange and correlation},\ }\href
  {https://doi.org/10.1103/PhysRevB.46.6671} {\bibfield  {journal} {\bibinfo
  {journal} {Phys. Rev. B}\ }\textbf {\bibinfo {volume} {46}},\ \bibinfo
  {pages} {6671} (\bibinfo {year} {1992})}\BibitemShut {NoStop}%
\bibitem [{\citenamefont {Perdew}\ \emph {et~al.}(1996)\citenamefont {Perdew},
  \citenamefont {Burke},\ and\ \citenamefont
  {Ernzerhof}}]{perdew1996generalized}%
  \BibitemOpen
  \bibfield  {author} {\bibinfo {author} {\bibfnamefont {J.~P.}\ \bibnamefont
  {Perdew}}, \bibinfo {author} {\bibfnamefont {K.}~\bibnamefont {Burke}},\ and\
  \bibinfo {author} {\bibfnamefont {M.}~\bibnamefont {Ernzerhof}},\ }\bibfield
  {title} {\bibinfo {title} {Generalized gradient approximation made simple},\
  }\href {https://doi.org/10.1103/PhysRevLett.77.3865} {\bibfield  {journal}
  {\bibinfo  {journal} {Phys. Rev. Lett.}\ }\textbf {\bibinfo {volume} {77}},\
  \bibinfo {pages} {3865} (\bibinfo {year} {1996})}\BibitemShut {NoStop}%
\bibitem [{\citenamefont {Dudarev}\ \emph {et~al.}(1998)\citenamefont
  {Dudarev}, \citenamefont {Botton}, \citenamefont {Savrasov}, \citenamefont
  {Humphreys},\ and\ \citenamefont {Sutton}}]{dudarev1998electronenergyloss}%
  \BibitemOpen
  \bibfield  {author} {\bibinfo {author} {\bibfnamefont {S.~L.}\ \bibnamefont
  {Dudarev}}, \bibinfo {author} {\bibfnamefont {G.~A.}\ \bibnamefont {Botton}},
  \bibinfo {author} {\bibfnamefont {S.~Y.}\ \bibnamefont {Savrasov}}, \bibinfo
  {author} {\bibfnamefont {C.~J.}\ \bibnamefont {Humphreys}},\ and\ \bibinfo
  {author} {\bibfnamefont {A.~P.}\ \bibnamefont {Sutton}},\ }\bibfield  {title}
  {\bibinfo {title} {Electron-energy-loss spectra and the structural stability
  of nickel oxide: An {LSDA+U} study},\ }\href
  {https://doi.org/10.1103/PhysRevB.57.1505} {\bibfield  {journal} {\bibinfo
  {journal} {Phys. Rev. B}\ }\textbf {\bibinfo {volume} {57}},\ \bibinfo
  {pages} {1505} (\bibinfo {year} {1998})}\BibitemShut {NoStop}%
\bibitem [{\citenamefont {Logvenov}\ \emph {et~al.}(2009)\citenamefont
  {Logvenov}, \citenamefont {Gozar},\ and\ \citenamefont
  {Bozovic}}]{logvenov2009hightemperature}%
  \BibitemOpen
  \bibfield  {author} {\bibinfo {author} {\bibfnamefont {G.}~\bibnamefont
  {Logvenov}}, \bibinfo {author} {\bibfnamefont {A.}~\bibnamefont {Gozar}},\
  and\ \bibinfo {author} {\bibfnamefont {I.}~\bibnamefont {Bozovic}},\
  }\bibfield  {title} {\bibinfo {title} {High-temperature superconductivity in
  a single copper-oxygen plane},\ }\href
  {https://doi.org/10.1126/science.1178863} {\bibfield  {journal} {\bibinfo
  {journal} {Science}\ }\textbf {\bibinfo {volume} {326}},\ \bibinfo {pages}
  {699} (\bibinfo {year} {2009})}\BibitemShut {NoStop}%
\end{thebibliography}%
	
\end{document}